\newif\ifANON{}
\newif\ifACM{}
\newif\ifLNCS{}
\newif\ifIACRTRANS{}
\newif\ifIEEE{}
\newif\ifUSENIX{}
\newif\ifELSEVIER{}
\newif\ifLINUXBUILD{}

\IEEEtrue{}

\IfFileExists{.tmp.anon.tmp}{\ANONtrue}{}
\IfFileExists{/dev/null}{\LINUXBUILDtrue}{}

\ifIEEE{}
\documentclass[conference]{IEEEtran}
\usepackage[table]{xcolor}
\usepackage[pdftex,colorlinks=true,pdfstartview=FitV,linkcolor=black,citecolor=blue,urlcolor=black]{hyperref}
\usepackage{placeins}
\usepackage{ulem}
\usepackage{booktabs}
\usepackage{graphicx}
\usepackage[shortcuts]{glossaries}
\newcommand{\appref}[1]{\hyperref[#1]{Appendix~\ref*{#1}}}
\glsdisablehyper{}

\newacronym{CSAM}{CSAM}{child sexual abuse material}

\usepackage[square,sort&compress,numbers]{natbib}

\ifANON{}
\author{Anonymous Submission}
\vspace{20pt}
\else
\author{

\author{
\IEEEauthorblockN{1\textsuperscript{st} Arttu Paju}
\IEEEauthorblockA{\textit{Tampere University} \\
Tampere, Finland \\
arttu.paju@tuni.fi}
\and
\IEEEauthorblockN{2\textsuperscript{nd} Waris Abdullah}
\IEEEauthorblockA{\textit{Tampere University} \\
Tampere, Finland \\
waris.abdullah@tuni.fi}
\and
\IEEEauthorblockN{3\textsuperscript{rd} Juha Nurmi}
\IEEEauthorblockA{\textit{Tampere University} \\
Tampere, Finland \\
juha.nurmi@tuni.fi}
}

}
\fi
\fi

\usepackage{xspace}
\usepackage{url} %
\usepackage{multirow}
\usepackage{subcaption}
\usepackage{listings}

\usepackage{CJKutf8} %

\hypersetup{
    colorlinks=true,
    linkcolor=blue,
    filecolor=magenta,
    urlcolor=cyan,
    citecolor=blue
}

\renewcommand{\sectionautorefname}{Section}

\newcounter{rqcounter}

\usepackage[inline]{enumitem}
\usepackage{xfp}        %
\usepackage{siunitx}    %
\newcommand{\rounding}[1]{%
  \num[round-mode=places, round-precision=2]{\fpeval{#1}}%
}

\makeatletter
\newcommand{\rqcounterautorefname}{RQ\@gobble}
\makeatother

\definecolor{roma1}{HTML}{7E1700} %
\definecolor{roma2}{HTML}{995215}
\definecolor{roma3}{HTML}{B0802B} %
\definecolor{roma4}{HTML}{C8B455} %
\definecolor{roma5}{HTML}{D0E3A3} %
\definecolor{roma6}{HTML}{A4E5D3} %
\definecolor{roma7}{HTML}{5DC1D3}
\definecolor{roma8}{HTML}{3292C2}
\definecolor{roma9}{HTML}{2064AE}
\definecolor{roma10}{HTML}{033198} %

\definecolor{light-gray}{gray}{0.95}

\newcommand{\qt}[1]{``#1''} %

\newcommand{\footurl}[1]{\footnote{\url{#1}}}

\newcommand{\TITLE}{Measuring onion website discovery and Tor users' interests with honeypots}

\newcommand{\CSAM}{\gls{CSAM}}
\newcommand{\ahmia}{Ahmia}
\newcommand{\pastebin}{pastebin.com}
\newcommand{\stronghold}{Stronghold paste}
\renewcommand{\sectionautorefname}{Section}

\ifLNCS{}

\fi
\ifIACRTRANS{}

\fi
\ifUSENIX{}

\fi
\ifACM{}

\fi
\ifIEEE{}
\makeatletter
\renewcommand{\@IEEEsectpunct}{~}
\makeatother

\fi

\newcommand{\KEYWORDS}{%
Tor,
Search engine,
Onion services,
Honeypot}

\title{\TITLE{}} %

\ifACM{}
\begin{abstract}
Tor enables anonymous web browsing and access to anonymous onion websites.
Prior work has focused on crawling and content analysis rather than on what users actually try to access.
Our honeypot approach measures engagement across onion-site categories, revealing behavioral interest rather than inferred popularity.
In March--April 2025, we deployed honeypot onion websites and seeded neutral-looking links via three channels---the \ahmia{} Tor search engine,
\stronghold{} onion \qt{paste} service, and \pastebin{}---to observe discovery and subsequent interaction events (CAPTCHA solves; registration/login attempts).
We observe that, almost without exception, human users originate from Ahmia.fi;
after removing the honeypot links from the Ahmia.fi search results, visits dropped to nearly zero and no users solved CAPTCHAs.
The honeypot landing front pages represent different forums for cybercrime
activities---child sexual abuse, violence, malware, stolen goods, illegal firearms, illegal drugs, and forgery items---and,
as a baseline comparison, an unclear forum.
Within that set, the CSAM-themed honeypot drew markedly higher engagement than the other honeypots.
When identical sites were offered in multiple languages, interaction events occurred most often on the English-language versions.
 \end{abstract}
\fi

\begin{document}

\ifIEEE{}
\IEEEoverridecommandlockouts{}
\makeatletter\def\@IEEEpubidpullup{6.5\baselineskip}\makeatother
\fi

\ifELSEVIER
\else
\maketitle
\fi

\ifLNCS{}
\begin{abstract}
Tor enables anonymous web browsing and access to anonymous onion websites.
Prior work has focused on crawling and content analysis rather than on what users actually try to access.
Our honeypot approach measures engagement across onion-site categories, revealing behavioral interest rather than inferred popularity.
In March--April 2025, we deployed honeypot onion websites and seeded neutral-looking links via three channels---the \ahmia{} Tor search engine,
\stronghold{} onion \qt{paste} service, and \pastebin{}---to observe discovery and subsequent interaction events (CAPTCHA solves; registration/login attempts).
We observe that, almost without exception, human users originate from Ahmia.fi;
after removing the honeypot links from the Ahmia.fi search results, visits dropped to nearly zero and no users solved CAPTCHAs.
The honeypot landing front pages represent different forums for cybercrime
activities---child sexual abuse, violence, malware, stolen goods, illegal firearms, illegal drugs, and forgery items---and,
as a baseline comparison, an unclear forum.
Within that set, the CSAM-themed honeypot drew markedly higher engagement than the other honeypots.
When identical sites were offered in multiple languages, interaction events occurred most often on the English-language versions.
  \and{} Honeypot
\end{abstract}
\fi

\ifIACRTRANS{}
\begin{abstract}
Tor enables anonymous web browsing and access to anonymous onion websites.
Prior work has focused on crawling and content analysis rather than on what users actually try to access.
Our honeypot approach measures engagement across onion-site categories, revealing behavioral interest rather than inferred popularity.
In March--April 2025, we deployed honeypot onion websites and seeded neutral-looking links via three channels---the \ahmia{} Tor search engine,
\stronghold{} onion \qt{paste} service, and \pastebin{}---to observe discovery and subsequent interaction events (CAPTCHA solves; registration/login attempts).
We observe that, almost without exception, human users originate from Ahmia.fi;
after removing the honeypot links from the Ahmia.fi search results, visits dropped to nearly zero and no users solved CAPTCHAs.
The honeypot landing front pages represent different forums for cybercrime
activities---child sexual abuse, violence, malware, stolen goods, illegal firearms, illegal drugs, and forgery items---and,
as a baseline comparison, an unclear forum.
Within that set, the CSAM-themed honeypot drew markedly higher engagement than the other honeypots.
When identical sites were offered in multiple languages, interaction events occurred most often on the English-language versions.
\end{abstract}
\fi

\ifIEEE{}
\begin{abstract}
Tor enables anonymous web browsing and access to anonymous onion websites.
Prior work has focused on crawling and content analysis rather than on what users actually try to access.
Our honeypot approach measures engagement across onion-site categories, revealing behavioral interest rather than inferred popularity.
In March--April 2025, we deployed honeypot onion websites and seeded neutral-looking links via three channels---the \ahmia{} Tor search engine,
\stronghold{} onion \qt{paste} service, and \pastebin{}---to observe discovery and subsequent interaction events (CAPTCHA solves; registration/login attempts).
We observe that, almost without exception, human users originate from Ahmia.fi;
after removing the honeypot links from the Ahmia.fi search results, visits dropped to nearly zero and no users solved CAPTCHAs.
The honeypot landing front pages represent different forums for cybercrime
activities---child sexual abuse, violence, malware, stolen goods, illegal firearms, illegal drugs, and forgery items---and,
as a baseline comparison, an unclear forum.
Within that set, the CSAM-themed honeypot drew markedly higher engagement than the other honeypots.
When identical sites were offered in multiple languages, interaction events occurred most often on the English-language versions.
 \end{abstract}

\fi

\ifUSENIX{}
\begin{abstract}
Tor enables anonymous web browsing and access to anonymous onion websites.
Prior work has focused on crawling and content analysis rather than on what users actually try to access.
Our honeypot approach measures engagement across onion-site categories, revealing behavioral interest rather than inferred popularity.
In March--April 2025, we deployed honeypot onion websites and seeded neutral-looking links via three channels---the \ahmia{} Tor search engine,
\stronghold{} onion \qt{paste} service, and \pastebin{}---to observe discovery and subsequent interaction events (CAPTCHA solves; registration/login attempts).
We observe that, almost without exception, human users originate from Ahmia.fi;
after removing the honeypot links from the Ahmia.fi search results, visits dropped to nearly zero and no users solved CAPTCHAs.
The honeypot landing front pages represent different forums for cybercrime
activities---child sexual abuse, violence, malware, stolen goods, illegal firearms, illegal drugs, and forgery items---and,
as a baseline comparison, an unclear forum.
Within that set, the CSAM-themed honeypot drew markedly higher engagement than the other honeypots.
When identical sites were offered in multiple languages, interaction events occurred most often on the English-language versions.
 \end{abstract}
\fi

\section{Introduction}\label{sec:intro}
Tor and Tor Browser are the most widely used tools for ensuring the anonymity of users and websites.
Tor enables anonymous hosting of websites using the \texttt{.onion} domain, whereas the Tor Browser enables anonymous web browsing and accessing these onion websites.
\ahmia{} (\texttt{Ahmia.fi}) Tor search engine is among the most common means by which individuals discover onion websites~\citep{DBLP:conf/uss/WinterERDCF18}.

Our research is user-oriented and based on their interests:
in March--April 2025, we deployed honeypot onion websites to see how human users try to create credentials for different types of onion websites.
Our onion honeypot websites appear to have legitimate cybercrime-related content, but they do not.
Instead, we monitor user activity to gather information about their behavior.
This research provides an overview of how users follow neutral entry descriptions and onion links from three different sources:
the \ahmia{} Tor search engine, the \stronghold{} onion \qt{paste} service, and \pastebin{}.
To detect humans and robots following the links, the websites are protected with CAPTCHAs.
This enables us to measure
(i) how users discover onion websites and
(ii) which honeypot onion websites human users access and try to create credentials on.
Our measurements are novel, as the related \textit{dark web} research focuses on web crawling rather than on studying how users access onion websites.
The following are our research questions (RQs).

\begin{description}
    \item[\textbf{\autoref{rq:find}}.]\refstepcounter{rqcounter}\label{rq:find}
    \textbf{Paste websites vs.\ Tor search engine: how do users find our honeypot onion websites?}
    \item[\textbf{\autoref{rq:access}}.]\refstepcounter{rqcounter}\label{rq:access}
    \textbf{What are the most captivating onion websites based on honeypot interaction data?}
    \item[\textbf{\autoref{rq:language}}.]\refstepcounter{rqcounter}\label{rq:language}
    \textbf{What languages do users prefer on our onion honeypot websites?}
\end{description}

Our contributions are as follows:
\begin{enumerate*}
\item\textbf{Discovery channels vs.\ human engagement:}
We shared links via \ahmia{}, \pastebin{}, and \stronghold{}.
While all three channels generated visits, most of human engagement (CAPTCHA solves and registration/log in attempts) in our dataset originated from \ahmia{}.

\item\textbf{Theme-level engagement:}
Within \ahmia{}-originated honeypot interactions, the \CSAM{}-themed honeypot showed disproportionately high engagement relative to other themes.
Surprisingly, illegal drugs was among the least engaged,
although related research on content available on Tor often concentrates on marketplaces selling illegal drugs.

\item\textbf{Language-level engagement:}
When we offered the same honeypot website in multiple domains and languages, English-language honeypots received the most interactions.
\end{enumerate*}

In \autoref{sec:related}, we outline the research gap that our findings address.
We provide an ethical rationale in \autoref{sec:ethics}.
In \autoref{sec:methods},
we explain our methods.
We present our results in \autoref{sec:results}.
The limitations of our work are listed in \autoref{sec:limitations}.
Finally, in \autoref{sec:conclusion}, we collect our findings and conclude our work.
In summary, our work provides the first quantitative measurement of how Tor search engines and paste services
lead users to attempt interaction with different categories of onion sites.
\section{Related work}\label{sec:related}

Tor routes traffic through multiple relays while encrypting the data between each relay
to ensure both the anonymity of the Tor user browsing websites and the anonymity of the websites available on Tor~\citep{DBLP:conf/uss/DingledineMS04}.
Tor onion websites (formerly known as \qt{hidden services}) constitute a substantial portion of the so-called \textit{dark web}.
There are roughly 800,000 onion domains~\citep{tormetrics}, but this does not reflect the number of onion websites,
as a website can have several onion domains,
and onion domains host all kinds of internet services, not only web services.

Existing research on the dark web has a focus on analyzing
content~\citep{aked2011investigation,DBLP:conf/isi/SpittersVS14,DBLP:conf/websci/ZabihimayvanSDA19,DBLP:journals/di/DalinsWC18,10.1145/3322645.3322691,DBLP:conf/IEEEares/BoshmafPKLJ23}.
Particularly, a significant portion of research concentrates on illegal marketplaces~\citep{DBLP:conf/www/Christin13,
DBLP:conf/uss/WegbergTSAGKCE18,
DBLP:conf/uss/SoskaC15,
DBLP:conf/IEEEares/NurmiNB23,
DBLP:journals/cacm/OosthoekCS23,
DBLP:conf/icdm/BaravalleLL16,
DBLP:conf/www/PortnoffADKBMLP17},
and identifying onion sites, but they lack insight into user interests.
While the popularity of various onion sites/categories---that
have been previously researched based on the content
available~\citep{DBLP:conf/www/TavabiBASFL19,
DBLP:conf/uss/CampobassoA23,
10.1145/3322645.3322691,
DBLP:journals/eswa/NabkiFAF19}---can
offer some insight into user interests,
it may not always accurately reflect user interest.

Our research is the first to directly quantify the types of onion websites that users actually try to log in to, i.e., not clicking and opening by accident or visited by a robot.
This has not been measured since 2014, when one study measured requests for resolving onion domain names,
but even then, this research method did not make a distinction between humans and automated crawlers accessing onion websites~\citep{DBLP:journals/iet-ifs/OwenS16}.
As such, the process for resolving onion domains contains automated crawlers accessing these websites and does not distinguish if users are actually eager to log in for the content.

While prior work offers limited insight into discovery mechanisms~\citep{DBLP:conf/soups/GallagherPM17},
most evidence comes from surveys~\citep{DBLP:conf/eurosp/Lusthaus19,zhang2020survey}, which introduce self-selection bias.
Based on a survey from August 16 to September 11, 2017, \ahmia{} is among the most common means by which individuals discover onion websites:
\qt{The three most popular ways that almost half of our survey participants discovered onion sites by were \dots (ii) search engines such as Ahmia, (46\%)}~\citep{DBLP:conf/uss/WinterERDCF18}.

In 2024, a study combined query logs from \ahmia{} and presented a survey targeting individuals who sought \CSAM{} on \ahmia{}
to investigate the number of users accessing this material and their motivations~\citep{nurmi2024investigating} finding that 11.1\% of users seek \CSAM{}\@.
Furthermore, this study estimated the number of \CSAM{} websites on Tor and found that one-fifth of the onion websites share \CSAM{}\@.

In another relevant study~\citep{DBLP:journals/chb/GalloSerpilloV24}, researchers deployed honeypots to geolocate IP addresses and capture real-time access patterns;
this highlighted regional variation on the dark web across different operating systems (e.g., \qt{Windows 10} and \qt{Android 10}).
Although the study offers useful information, it does not clarify how users find these honeypot websites or the popularity between categories.

Research into financial transaction statistics on the dark web,
specifically investigating which category generated the most revenue,
shows that sexual abuse generated the highest revenue (94,241,807 US dollars) in 2021,
followed by drugs, cybercrime, and extremism~\citep{DBLP:journals/access/OosthoekSS23}.

An article published in 2020~\citep{DBLP:conf/iwcmc/ZeidMB20} implemented two secure honeypots---a chatroom and a website.
Over a period of seven months, the study discovered that \CSAM{} accounted for 50\% of chatroom requests, with discussions on pornography and video links following closely behind.
The honeypots also faced attacks, including scripting and SQL injections.
\section{Ethics}\label{sec:ethics}

This research complies with relevant laws, regulations, and guidelines.
The Ahmia Terms of Service include a statement that Ahmia is used for scientific research and carries out tests with the interface.
Particularly, Ahmia's Terms of Service mention A/B tests and edited result views and the goal to carry out tests with different layouts, messages,
or redirect options on a randomized basis to understand user preferences to generate scientific knowledge and benefit society.
To protect users' privacy, the Ahmia search engine does not record their IP addresses, does not use cookies, and only saves minimal HTTP logs.
Similarly, honeypot onion websites only save minimal HTTP logs, and users remain anonymous as they access them via Tor.
We did not track users with cookies because doing so requires users' permission, which we did not want to add as an extra step.
We do not compromise the users' anonymity and privacy in any way.
We do not cause any harm to any users.
\section{Methods}\label{sec:methods}
We shared honeypot URLs on three sources---(i) the \ahmia{} Tor search engine;
(ii) the \stronghold{} onion service; and
(iii) \pastebin{}---each with a unique onion domain to separate discovery origins.

Each of our honeypot websites included two pages.
A neutral-looking landing page with a CAPTCHA puzzle,
and a second page---that could be accessed after solving the CAPTCHA---with
descriptions of some type of cybercrime activity as well as registration/login fields.
The registration and login fields on these pages had no real functionality,
but they did leave a trace in logs that tracked the number of users who attempted to create credentials or log in to the sites.
We made no distinction between how many users attempted to create credentials or log in to the sites.
Instead, we viewed both actions as attempts to access the content available on the honeypot forum.

Our data collection period started on March 24, 2025 by publishing the honeypot websites and distributing the URLs.
The URL distribution happened by randomly showing URLs to the honeypot sites alongside other search results on \ahmia{}
and by publishing the URLs on both \pastebin{} and \stronghold{}.
We only showed the URLs on \pastebin{} and \stronghold{} for the duration of one day on March 24.
On April 10, 2025, we removed the URLs from \ahmia{}.
Thereafter, we continued to monitor the activity on all URLs for an additional week.
Thus, the total analysis period ran from March 24 to April 17, 2025.

We modified the search engine to randomly add a neutral description of \qt{onion log in page available} to the search results,
along with a random link when there were more than 10 search results.
As a comparison, in March 2024, \ahmia{} recorded an average of 137,993 daily searches, equivalent to 95 searches per minute.
Furthermore, \ahmia{} filters \CSAM{} from the search results and prevents \CSAM{}-related queries.

To quantify how users find onion services, we measured the number of visits to our honeypot websites from each source.
This allowed us to compare the popularity of each source in accessing onion services and answer the~\autoref{rq:find}.

We used CAPTCHA puzzles to protect the onion websites
and distinguish real human users from bots and crawlers that follow URLs on the Tor network.
Realistically, only few crawlers/bots are capable of solving CAPTCHAs.
This is particularly true when the website is brand new and no human has instructed the crawler on how to solve this new CAPTCHA type.
Thus, the number of solved CAPTCHAs indicates the number of real human users
who attempted to interact with the honeypot sites.

\begin{figure}[!htbp]
  \centering
    \includegraphics[width=\linewidth]{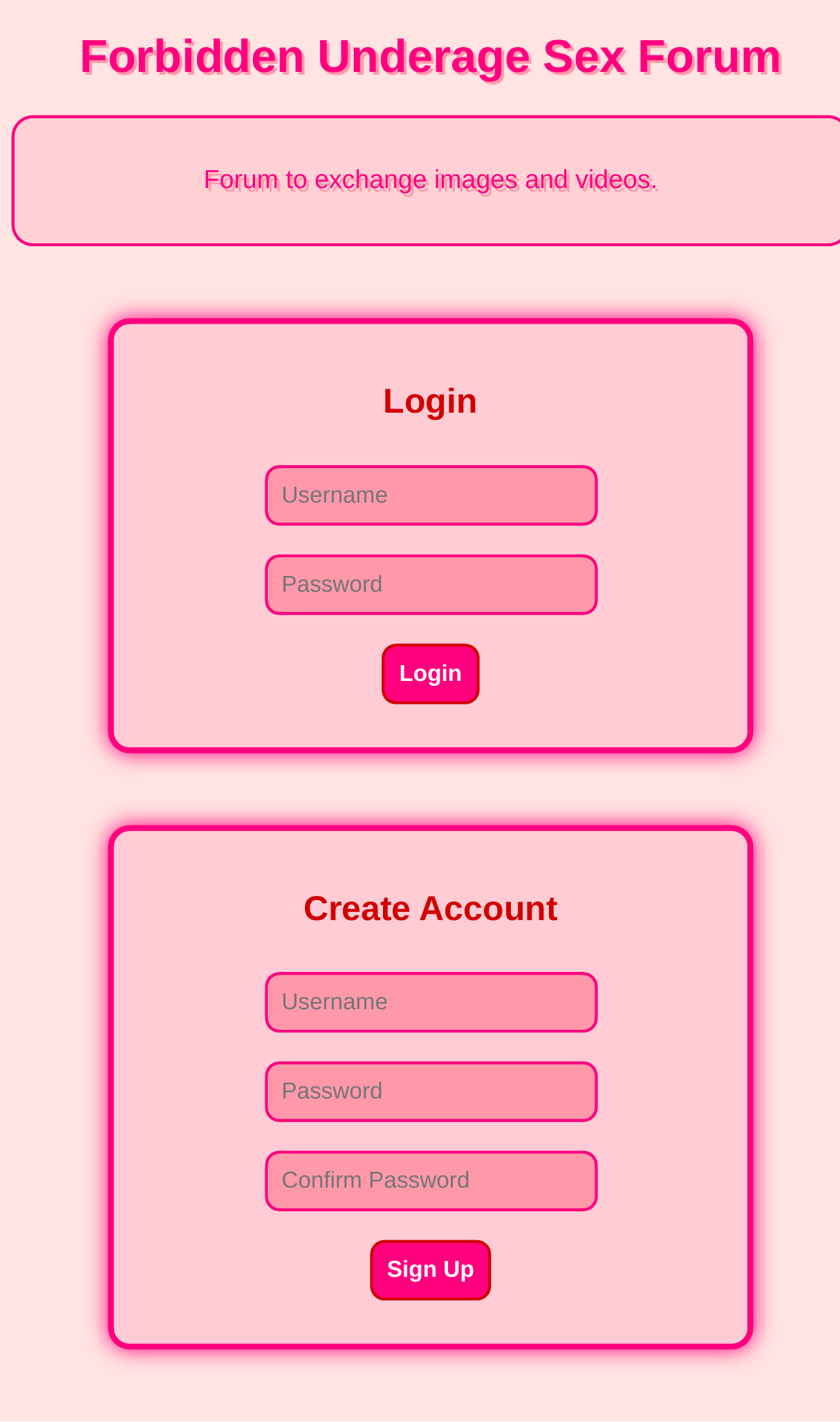}
    \caption{\CSAM{} honeypot (English version). All honeypot websites contained only text-based hints for users,
    along with free-to-use legal images, and did not contain any illegal content.}\label{fig:honeypots}
\end{figure}

To quantify what kind of onion websites the users try to access,
we created honeypot websites representing eight different categories:
(i) \CSAM{};
(ii) violence;
(iii) malware;
(iv) stolen goods;
(v) illegal firearms;
(vi) illegal drugs; and
(vii) forgery items.
We selected these categories because prior content-analysis studies show that they are
among the most prevalent categories of illegal activities on Tor network~\citep{DBLP:conf/eacl/NabkiFAP17,10.1145/3322645.3322691}.
Additionally, we created a generic (viii) unclear forum front page that acted as a baseline comparison.
As an example,~\autoref{fig:honeypots} shows the honeypot website relating to the
\CSAM{} in the English language that we used in the research.
Screenshots of other honeypot websites in the English language are in the~\appref{app:appendix}.

By analyzing the number of attempts to create credentials or log in on these sites related to different categories,
we can determine which websites the users are actually interested in and trying to access,
thus answering the~\autoref{rq:access}.

To quantify the disparity between users' preferences for language variants,
we implemented identical websites for each category in four languages:
English,
German,
Finnish, and
Russian.
We selected English, German, Finnish, and Russian since prior studies identify them as prevalent onion-site languages~\citep{DBLP:conf/isi/SpittersVS14,10.1145/3322645.3322691}.
By comparing the number of attempts to create credentials or log in on different language sites,
we can determine the popularity of each language among users.
This allowed us to answer the~\autoref{rq:language}.

In total, we created honeypot sites for eight categories in four languages, resulting in 32 unique websites.
We also created separate domains for each of the three sources, which resulted in 96 unique URLs for the research.

The unit of analysis in our research is honeypot interaction events.
Tor's anonymity and our minimal logging prevent reliable de-duplication of individuals;
a single person may generate multiple events (e.g., multiple registration attempts after one CAPTCHA).
Accordingly, any references to \qt{user interest} should be interpreted as aggregate event-level engagement, not person-level preferences.
\section{Results}\label{sec:results}
We tracked the number of visits, number of solved CAPTCHAs, and attempts to register or log in
to our honeypot sites from each source (\ahmia{}, \pastebin, and \stronghold{}).
Comparing the number of visits and solved CAPTCHAs originating from each source allows us to answer the~\autoref{rq:find},
and comparing the number of attempts to register/log in for different honeypot sites
allows us to answer the~\autoref{rq:access} and the~\autoref{rq:language}.

\subsection{RQ1\@: Paste websites vs.\ Tor search engine: how do users find our honeypot onion websites?}
\begin{figure}[!htb]
    \includegraphics[width=0.99\linewidth]{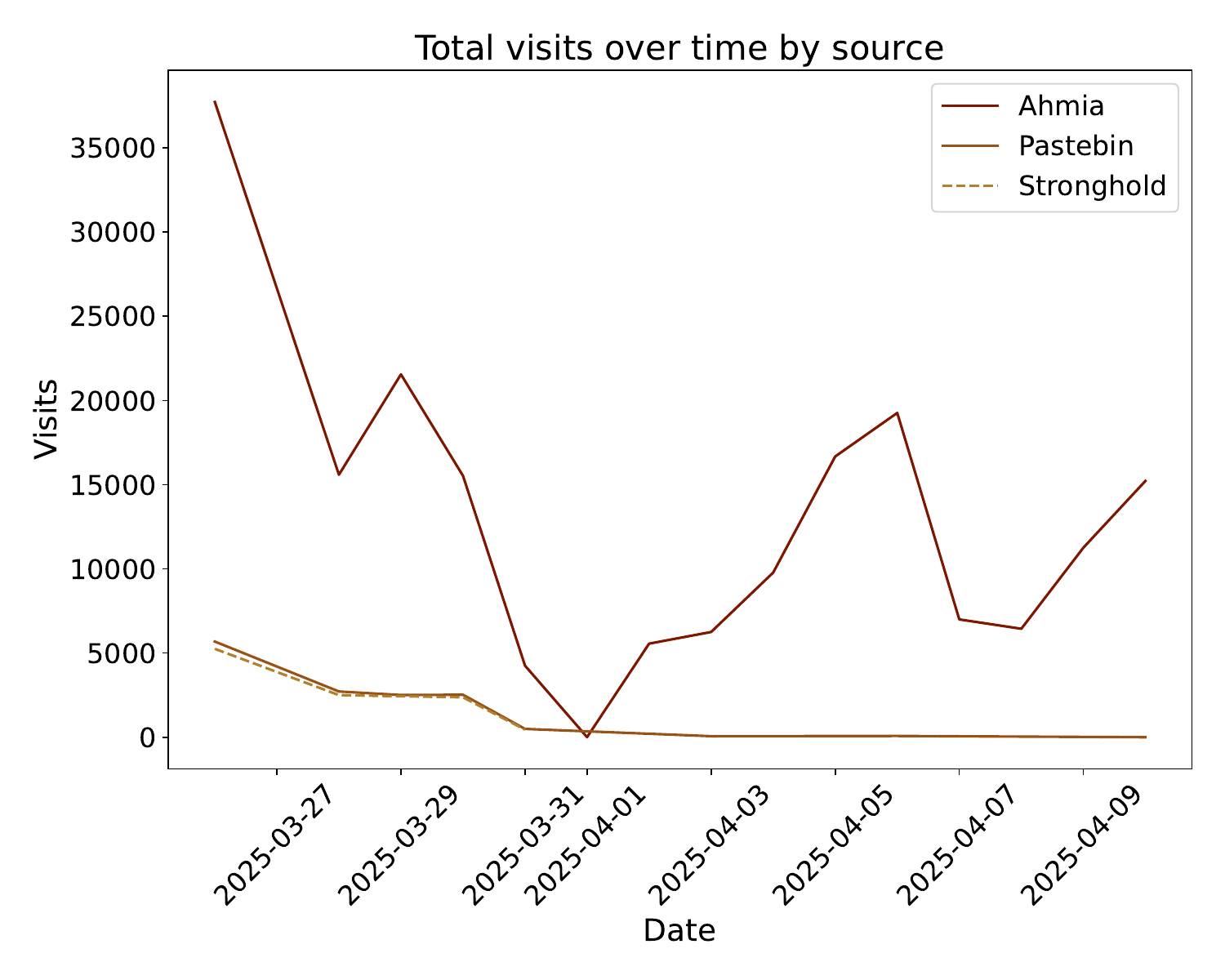}
    \caption{
      Visits to our honeypot websites from all three distribution sources.
    }\label{fig:hitsOverTime}
\end{figure}

Between March 26, 2025 and April 10, 2025---while distributing links to our honeypot sites---we
received a combined total of 219,173 visits on all honeypot sites,
with \rounding{192101/219173*100}\,\% (n = 192,101) coming from \ahmia{}, \rounding{14066/219173*100}\,\% (n = 14,066) from \pastebin{},
and \rounding{13006/219173*100}\,\% (n = 13,006) from \stronghold{}, respectively.
The total number of visits over time for each source is presented in~\autoref{fig:hitsOverTime}.

During the first day of the trial, the majority of visits originated from the \ahmia{} search engine,
while \pastebin{} and \stronghold{} received less traffic by an order of magnitude.
The difference in the number of visits from \pastebin{} and \stronghold{} was negligible.
On the first day of the analysis (March 26), we had significantly more visits from all three sources than on any other day.
This is to be expected, as whenever new onion sites are indexed on Tor, active crawlers send massive amounts of traffic to those sites.
After the first day, we received fluctuating visits from \ahmia{},
but at least thousands of daily visits throughout the analysis period,
except on April 1 when the search functionality was not working properly.

\begin{figure}[!htb]
    \includegraphics[width=0.99\linewidth]{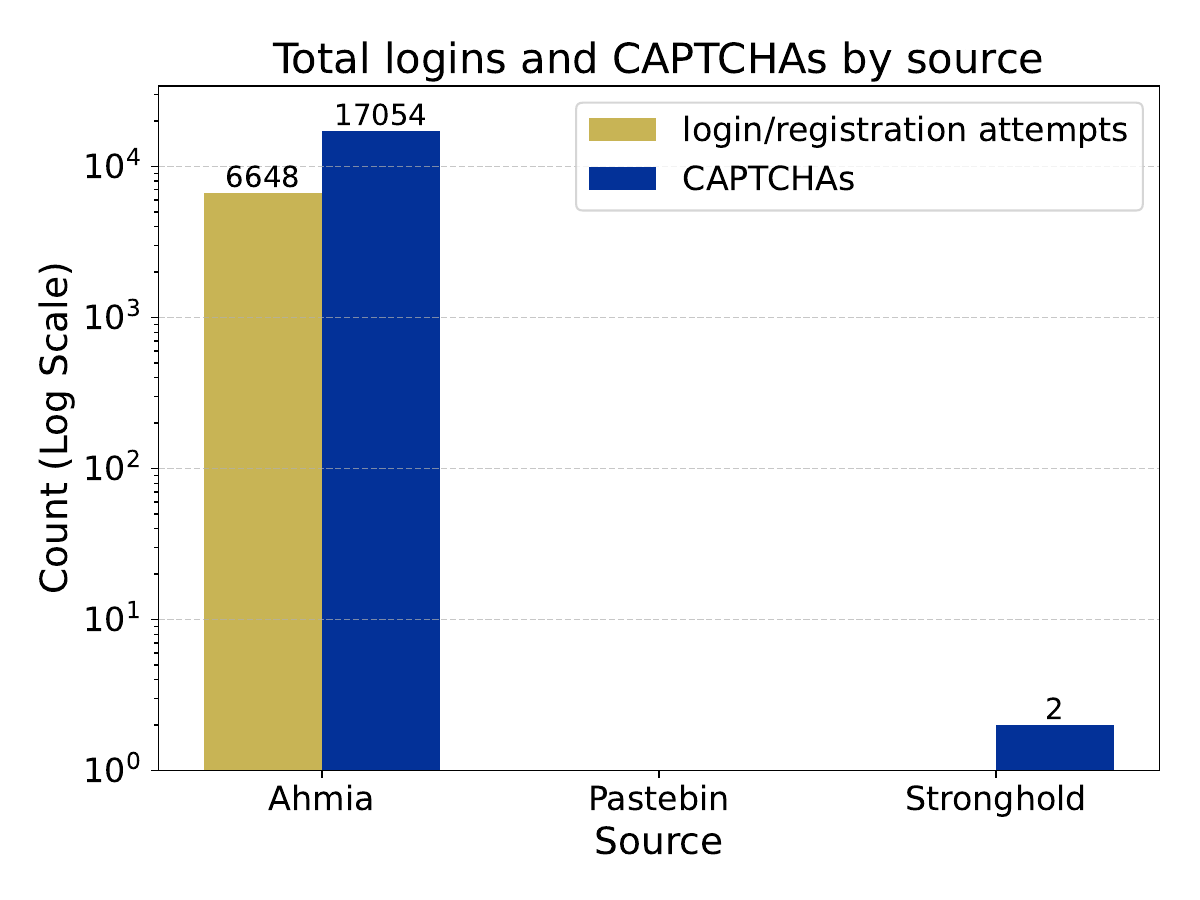}
    \centering
    \caption{
      Resolved CAPTCHAs and log in attempts from all three sources.
    }\label{fig:captchasVsLogins}
\end{figure}

Two \qt{paste} websites, \pastebin{} and \stronghold{}, displayed the links for 24 hours to compare their popularity on the first trial day.
Please note that these paste websites only highlight the most recently added content, and \stronghold{} even automatically removes the paste after 24 hours.
Thus, we compare the popularity of Ahmia users and paste site users on the first day.
We see some visiting activity after the first day, which reveals that some automated systems collected the onion links from these paste websites and continued to visit them.
Nevertheless, these visits are not mainly from human users, as only two of them actually led to resolving the CAPTCHA,
and there have been no log in attempts; we will discuss these findings further next.
During our active URL sharing period between March 26, 2025 and April 10, 2025,
our dataset recorded a combined total of 17,056 solved CAPTCHAs (\rounding{17056/219173*100}\,\% of visits leading to solved CAPTCHAs)
and 6,648 attempts to register or log in (\rounding{6648/219173*100}\,\% of visits leading to registration/log in attempts).
The total count of solved CAPTCHAs and attempts to register/log in originating from each source is presented in~\autoref{fig:captchasVsLogins}.

We observed a continuous stream of solved CAPTCHAs and log in attempts exclusively from \ahmia{}'s search results.
In total, 17,054 solved CAPTCHAs and 6,648 registration/log in attempts originated from \ahmia{}.
Only two completed CAPTCHAs originated from \stronghold{}.
No registration/log in attempts originated from either \pastebin{} or \stronghold{}.
\textit{These results show that real users prefer \ahmia{} search engine over paste sites for accessing dark web onion websites}.
\stronghold{} and \pastebin{} do not represent actual user traffic, but rather bots and crawlers.

\begin{figure}[!htb]
  \centering
  \includegraphics[width=\linewidth]{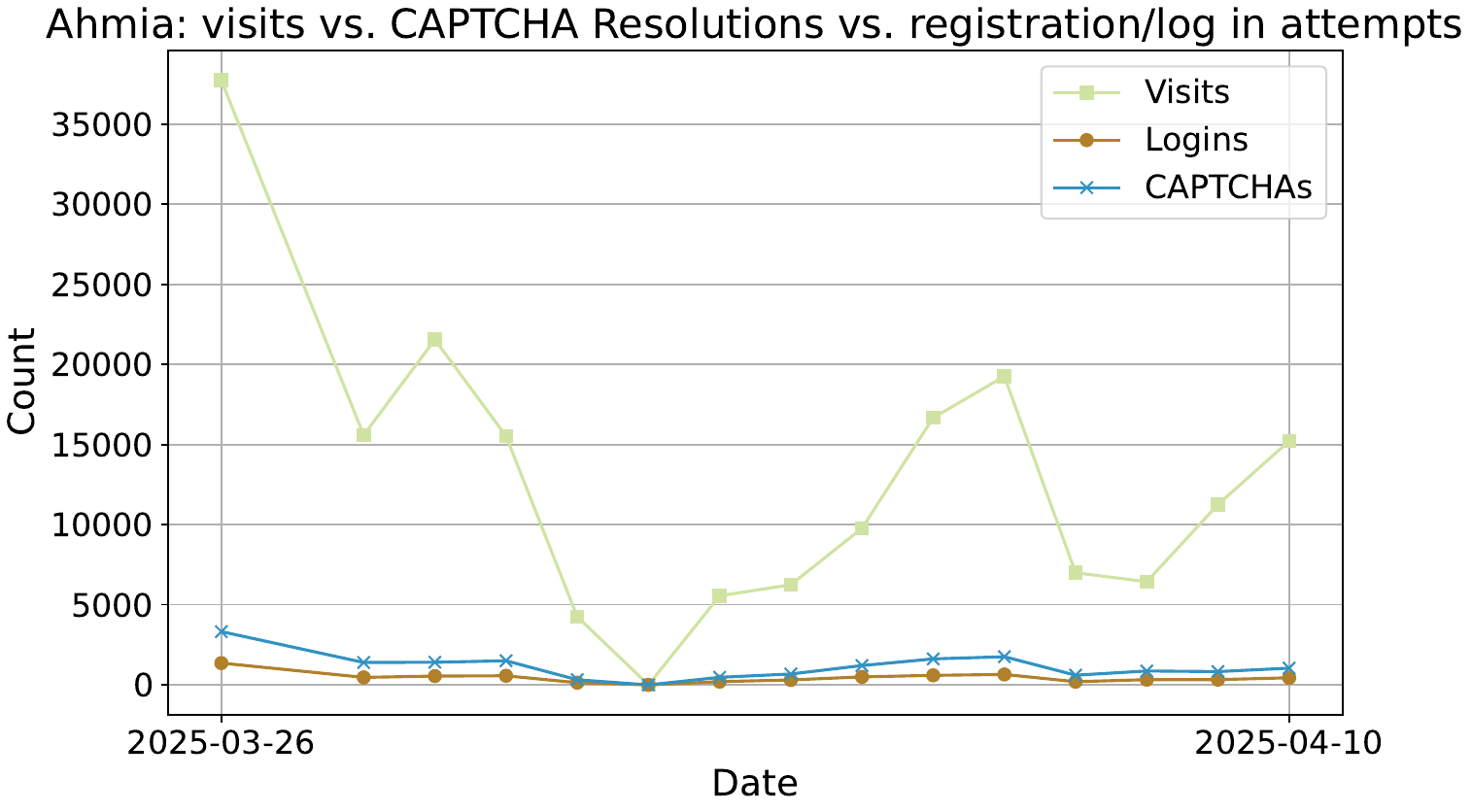}
  \caption{
  Visits, solved CAPTCHAs, and registration/log in attempts originating from \ahmia{}'s results.
  }\label{fig:attemps}
\end{figure}

\autoref{fig:attemps} shows the total number of visits, solved CAPTCHAs, and registration/log in attempts from \ahmia{}.
The solved CAPTCHAs and registration/log in attempts follow the trend of overall visits quite consistently.
On March 26---the first day of the analysis---there were 3,329 CAPTCHAs solved and 1,362 registration/log in attempts,
which correspond to a \rounding{3329/48663*100}\,\% CAPTCHA-solving rate and a \rounding{6648/17056*100}\,\% follow-up registration rate among those who solved the CAPTCHA\@.
These figures had dropped to 1,045 and 444, respectively, by the end of the URL sharing period on April 10.
Only a small fraction of the visits resulted in CAPTCHA resolutions,
and an even smaller number led to registration/log in attempts.
After removing the URLs from \ahmia{} on April 10, we continued to monitor the activity on all URLs for an additional week.
We did not detect any attempts to create credentials or log in.
In fact, the CAPTCHA was solved only 20 times, and there were only 1,053 visits during this seven-day period.
This again implies that \ahmia{} was the primary source for human users accessing the onion addresses during the investigation.

\subsection{RQ2\@: What are the most captivating onion websites based on honeypot interaction data?}
\begin{figure}[!htb]
  \centering
  \begin{subfigure}[b]{0.49\textwidth}
    \centering
    \includegraphics[width=\linewidth]{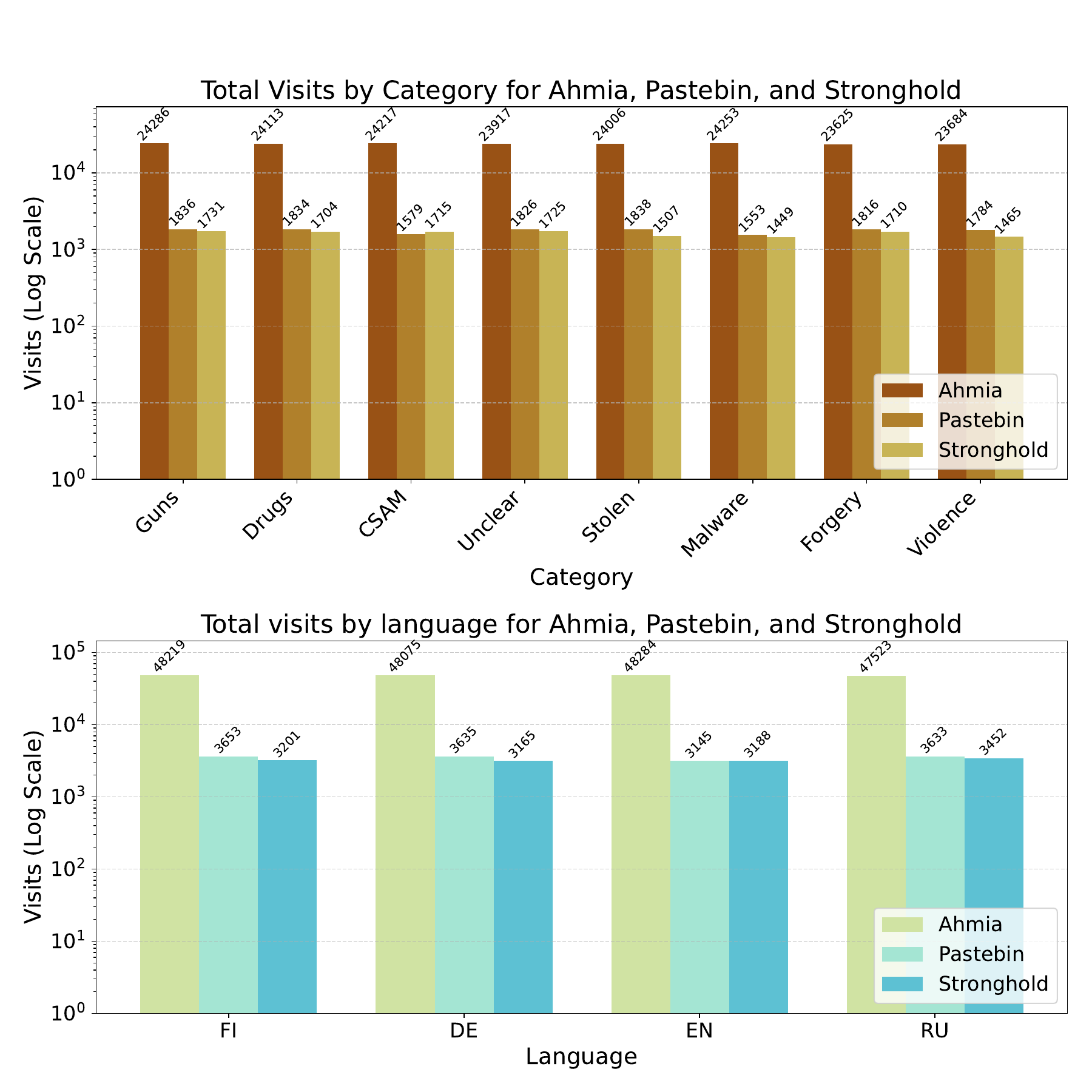}
    \caption{
    Total visits by category and language for \ahmia{}, \pastebin{}, and \stronghold{}.
    }\label{fig:combinedVisitsByCategoryAndLanguage}
  \end{subfigure}%
  \hfill
  \begin{subfigure}[b]{0.49\textwidth}
    \centering
    \includegraphics[width=\linewidth]{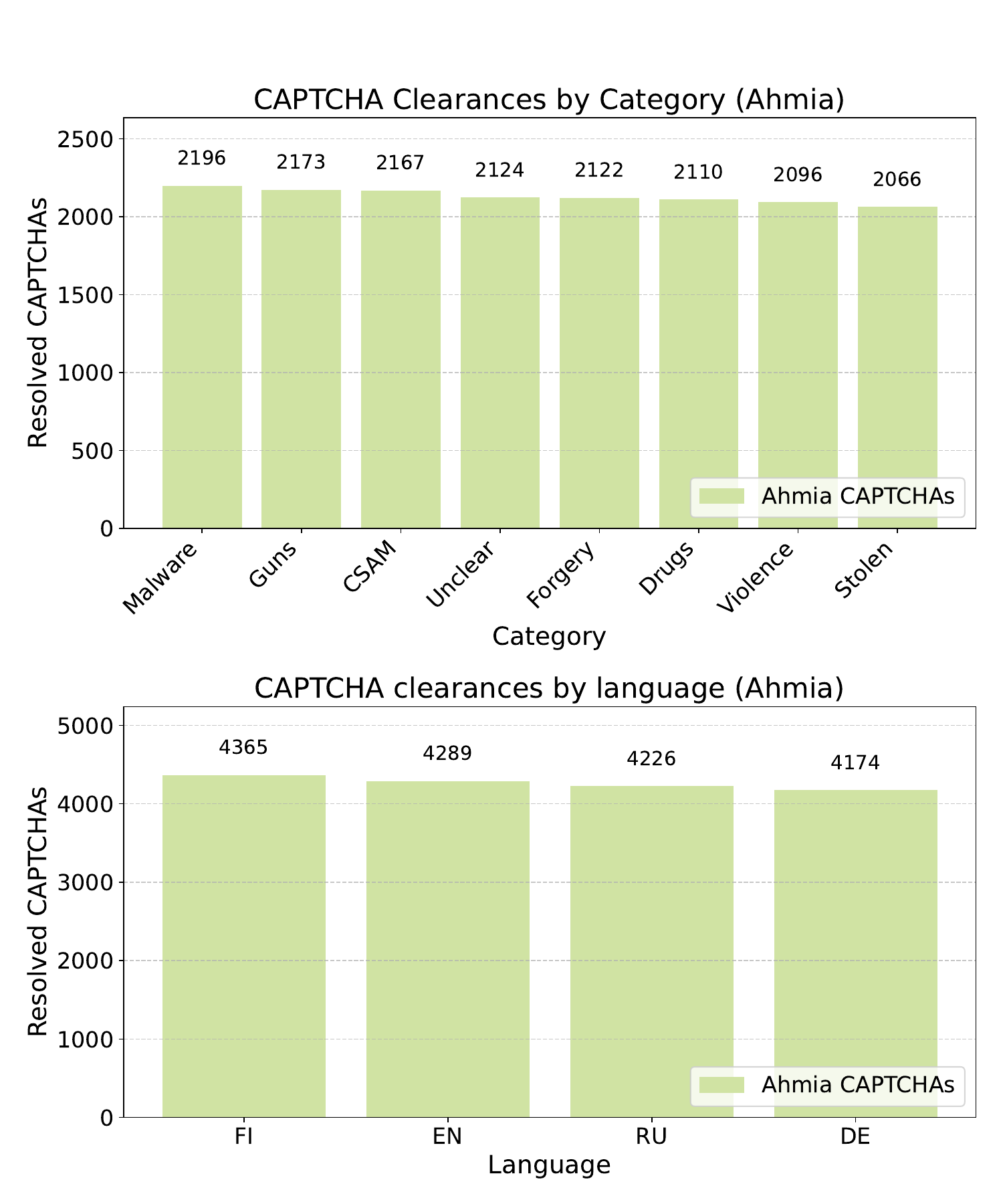}
    \caption{
    Resolved CAPTCHAs for each category and language of websites for traffic coming from \ahmia{}.
    }\label{fig:combinedCaptchaClearenceByCategoryAndLanguage}
    \end{subfigure}%
  \caption{
  All honeypot websites---as expected---received roughly equal amount of visits and resolved CAPTCHAs.}\label{fig:uniformity}
\end{figure}

We analyzed traffic to our honeypot sites in eight distinct categories
(\CSAM{}, violence, malware, stolen goods, illegal firearms, illegal drugs, forgery items, and unclear)
and four distinct languages (English, German, Finnish, and Russian).
Different categories and languages have different numbers of registration/log in attempts, indicating user preferences.
Thus, comparing the total number of registration/log in attempts on different categories and languages
allows us to determine the type of content users seek on the dark web as well as their language preferences.
Because \ahmia{} displays a neutral \qt{Onion login page}
and URL for all websites without content information before CAPTCHA clearance,
and URLs posted in \pastebin{} and \stronghold{} do not show the context of honeypot sites,
all categories and languages receive equal traffic and CAPTCHA clearances,
as expected.
\autoref{fig:combinedVisitsByCategoryAndLanguage} shows a consistent distribution of visits across categories and languages,
while~\autoref{fig:combinedCaptchaClearenceByCategoryAndLanguage} shows
a similar consistent distribution of CAPTCHA clearances across categories and languages.
Thus, the number of visits or resolved CAPTCHAs cannot be used for
determining the user preference for different categories or languages.
However, the number of registration/log in attempts indicates genuine user interest,
as only authentic users typically bypass the CAPTCHA and gain access to the site's content,
and users will only attempt to create credentials or log in if they are interested in the material.

\begin{figure}[!htb]
  \centering
  \includegraphics[width=\linewidth]{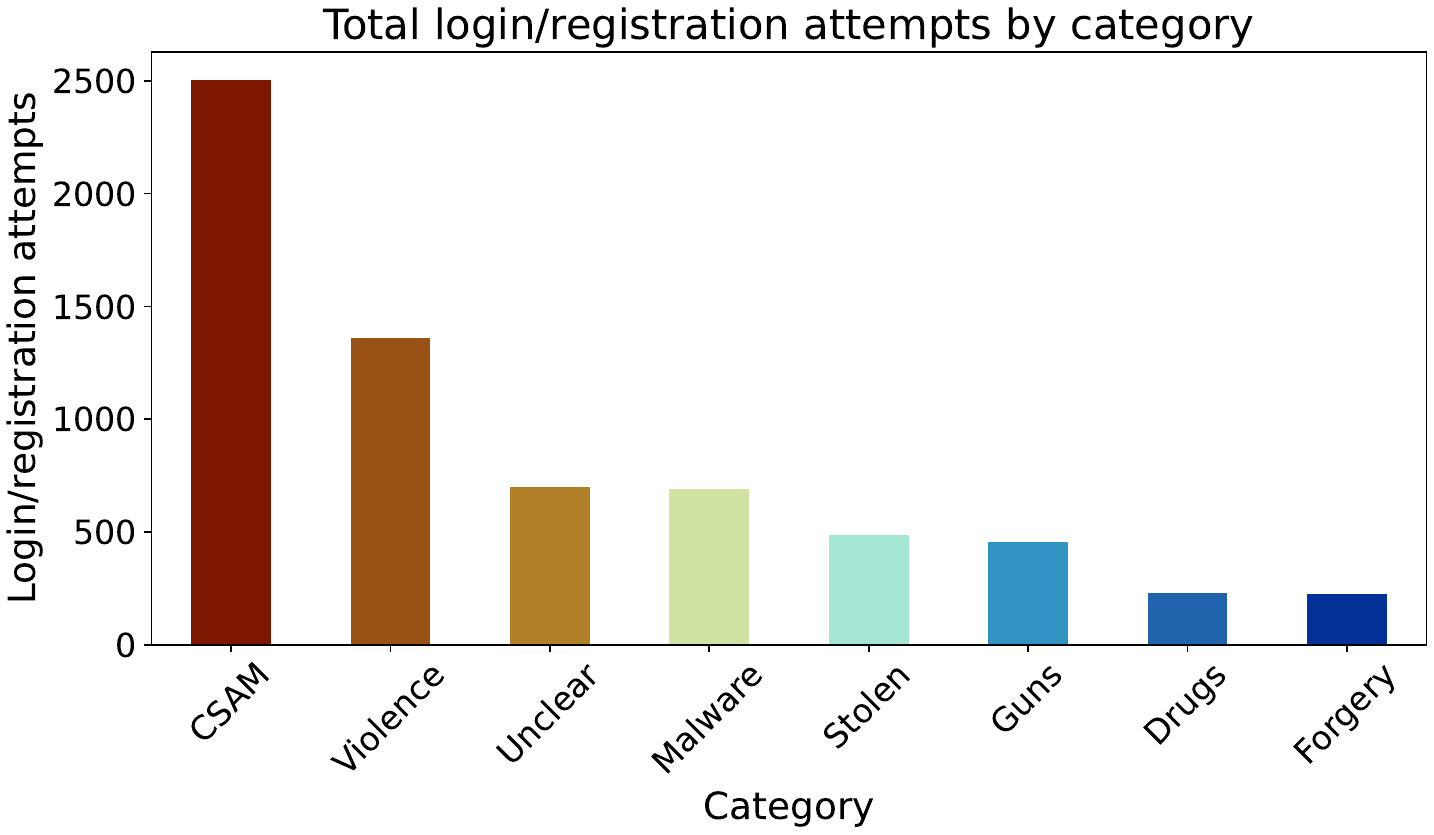}
  \caption{
  Total registration/log in attempts for each category of websites.
  }\label{fig:category}
\end{figure}

We recorded a total of 6,648 registration/log in attempts originating from \ahmia{}.
\autoref{fig:category} displays the number of attempts for each category.
\CSAM{} was by far the most popular category, with \rounding{2504/6648*100}\,\% (n = 2,504) of all registration/log in attempts.
Violence was the second most popular category by a significant margin, with \rounding{1361/6648*100}\,\% (n = 1,361) of all attempts.
Following these, the baseline unclear and malware categories received roughly equal interest,
with \rounding{699/6648*100}\,\% (n = 699) and \rounding{689/6648*100}\,\% (n = 689) of all attempts, respectively.
Following that, the stolen goods and illegal firearms categories also saw roughly equal popularity,
with \rounding{486/6648*100}\,\% (n = 486) and \rounding{456/6648*100}\,\% (n = 456) of all attempts, respectively.
Surprisingly, illegal drugs was the second least popular category, with only \rounding{228/6648*100}\,\% (n = 228) of all attempts.
The least popular category was forgery items, which had \rounding{225/6648*100}\,\% (n = 225) of all attempts---just barely less than illegal drugs.
There were no registration/log in attempts from \pastebin{} or \stronghold{},
again indicating that the majority of visits from these sources were bots or crawlers.

The statistics in~\autoref{fig:captcha_login_category} show the percentage of users
who tried to register/log in to each category of websites after resolving the CAPTCHAs.
We only included the traffic from \ahmia{},
since no attempts to register/log in originated from either \stronghold{} or \pastebin{}.

\begin{table}[!htb]
  \centering
  \footnotesize
  \caption{
    Percentage of registration/log in attempts per resolved CAPTCHA for each category.
  }\label{fig:captcha_login_category}
  \begin{tabular}{
      l
      >{\centering\arraybackslash}p{2.0cm}
      >{\centering\arraybackslash}p{2.0cm}
      >{\centering\arraybackslash}p{1.6cm}
  }
      \toprule
      \textbf{Category} & \textbf{CAPTCHAs} & \textbf{Log./cred.creat.\ attempts} & \textbf{Percentage} \\
      \midrule
      \CSAM{}           & 2,167 & 2,504 & 115.55 \\
      Violence          & 2,096 & 1,361 & 64.93 \\
      Unclear           & 2,124 & 699   & 32.91 \\
      Malware           & 2,196 & 689   & 31.38 \\
      Stolen goods      & 2,066 & 486   & 23.52 \\
      Illegal firearms  & 2,173 & 456   & 20.98 \\
      Illegal drugs     & 2,110 & 228   & 10.81 \\
      Forgery           & 2,122 & 225   & 10.60 \\
      \midrule
      \textbf{Total} & 17,054 & 6,648 & 38.98 \\
      \bottomrule
  \end{tabular}
\end{table}

After resolving the CAPTCHA, an average of \SI{38.98}{\percent} of users attempted to register/log in to the website.
This number varied greatly depending on the category.
The \CSAM{} category had an unusually high rate of user engagement at \SI{115.55}{\percent},
indicating multiple registration/log in attempts per user who solved the CAPTCHA\@,
underscoring that results are event-level rather than person-level.
The violence category also had significant user engagement at \SI{64.93}{\percent}.
On the other hand, illegal drugs and forgery had significantly lower user engagement rates
than the average at \SI{10.81}{\percent} and \SI{10.60}{\percent}, respectively.
Overall---as expected---the user engagement rate closely matches
the number of registration/log in attempts per website category (see~\autoref{fig:category}).
\textit{These interactions show disproportionately high user engagement with the \CSAM{}- and violence-themed honeypots}.
The unexpectedly high \CSAM{} engagement---despite Ahmia's filtering---may indicate that
search-engine-referred visitors are a less experienced subset of Tor users who engage with first-accessible content.

\subsection{RQ3\@: What languages do users prefer on our onion honeypot websites?}
\begin{figure}[!htb]
    \centering
        \includegraphics[width=\linewidth]{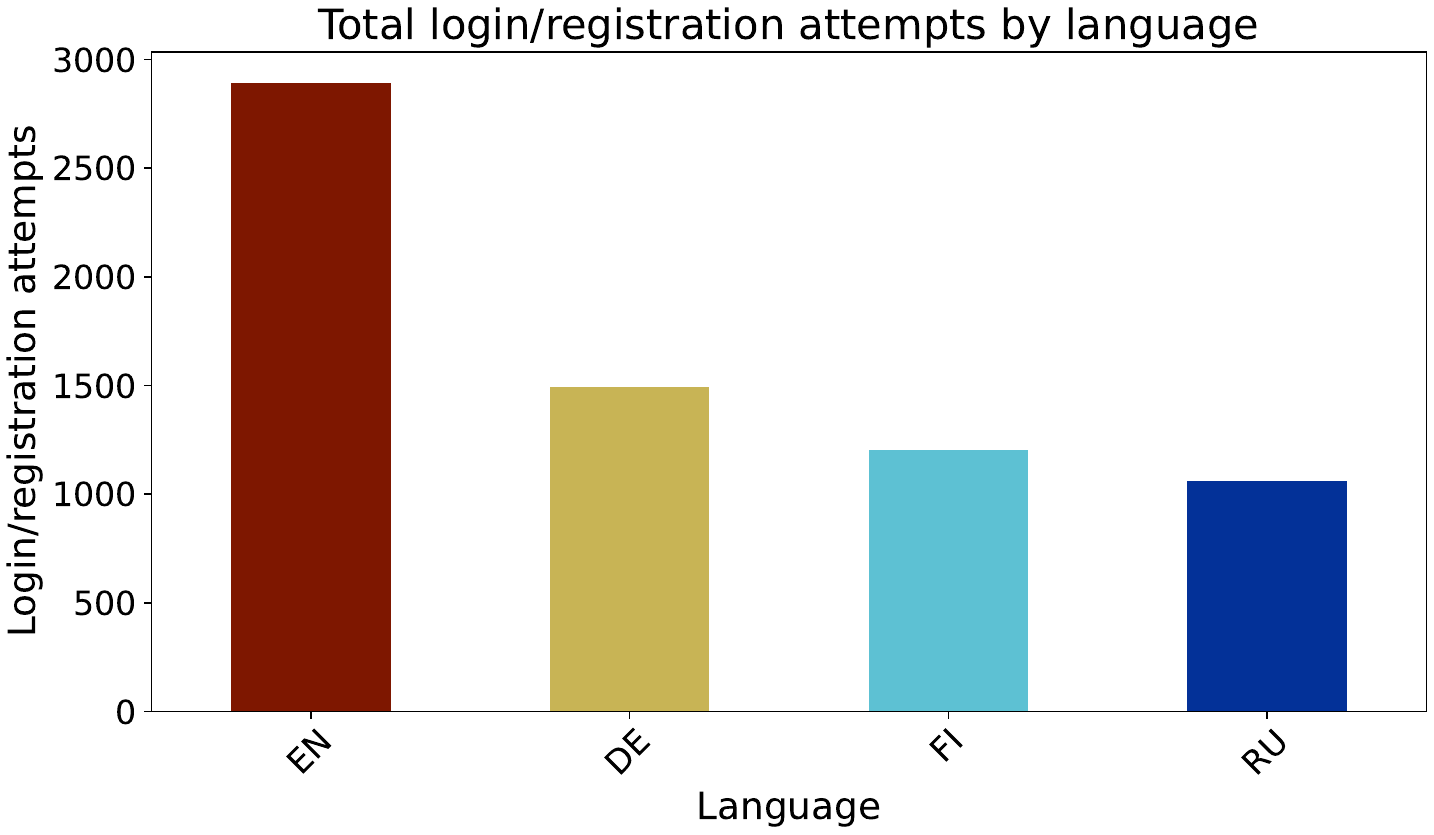}
        \caption{Through all categories, the log in attempts by language.}\label{fig:all_language}
\end{figure}

When the same category of a website is available in multiple languages, the number of registration/log in attempts reveals the popularity between them.
\autoref{fig:all_language} displays the number of registration/log in attempts per language.
The most popular language was English, with \rounding{2891/6648*100}\,\% (n = 2,891 of 6,648) of all  attempts,
followed by German and Finnish (\rounding{1494/6648*100}\,\% (n = 1,494) and \rounding{1201/6648*100}\,\% (n = 1,201) of all attempts, respectively).
Russian was the least-popular language with \rounding{1062/6648*100}\,\% (n = 1,062) of all attempts.
The percentage of resolved CAPTCHAs leading to log in attempts is detailed in~\autoref{fig:captcha_login_language}.
English shows the highest conversion rate at \SI{67.40}{\percent},
followed by German at \SI{35.8}{\percent},
Finnish at \SI{27.5}{\percent},
and Russian at \SI{25.1}{\percent}.

\begin{table}[!htb]
  \centering
  \caption{Percentage of log in attempts per resolved CAPTCHAs by language on \ahmia{}.\label{fig:captcha_login_language}}
  \begin{tabular}{
      >{\arraybackslash}p{2.0cm}
      >{\centering\arraybackslash}p{2.0cm}
      >{\centering\arraybackslash}p{2.0cm}
      >{\centering\arraybackslash}p{1.6cm}
  }
      \toprule
      \textbf{Language} & \textbf{CAPTCHAs} & \textbf{Log./cred.creat.\ attempts} & \textbf{Percentage} \\
      \midrule
      English & 4,289 & 2,891 & 67.40 \\
      German  & 4,174 & 1,494 & 35.79 \\
      Finnish & 4,365 & 1,201 & 27.51 \\
      Russian & 4,226 & 1,062 & 25.13 \\
      \midrule
      \textbf{Total} & 17,054 & 6,648 & 38.98 \\
      \bottomrule
  \end{tabular}
\end{table}

Despite being generally more prominent on onion websites~\citep{DBLP:conf/isi/SpittersVS14,10.1145/3322645.3322691}
and having more global speakers~\citep{Ethnologue},
Russian was less popular in our dataset than Finnish.
The Finnish language overrepresentation in our dataset might be explained
by the fact that \ahmia{} is a Finnish search engine,
so it may have a higher popularity among the Finnish-speaking population compared to other websites.
Similarly, Russian Tor users may not use \ahmia{} as frequently as English-, German-, or Finnish-speaking users,
thus resulting in underrepresentation of the language in our dataset.

\begin{figure}[!htb]
    \centering
    \includegraphics[width=\linewidth]{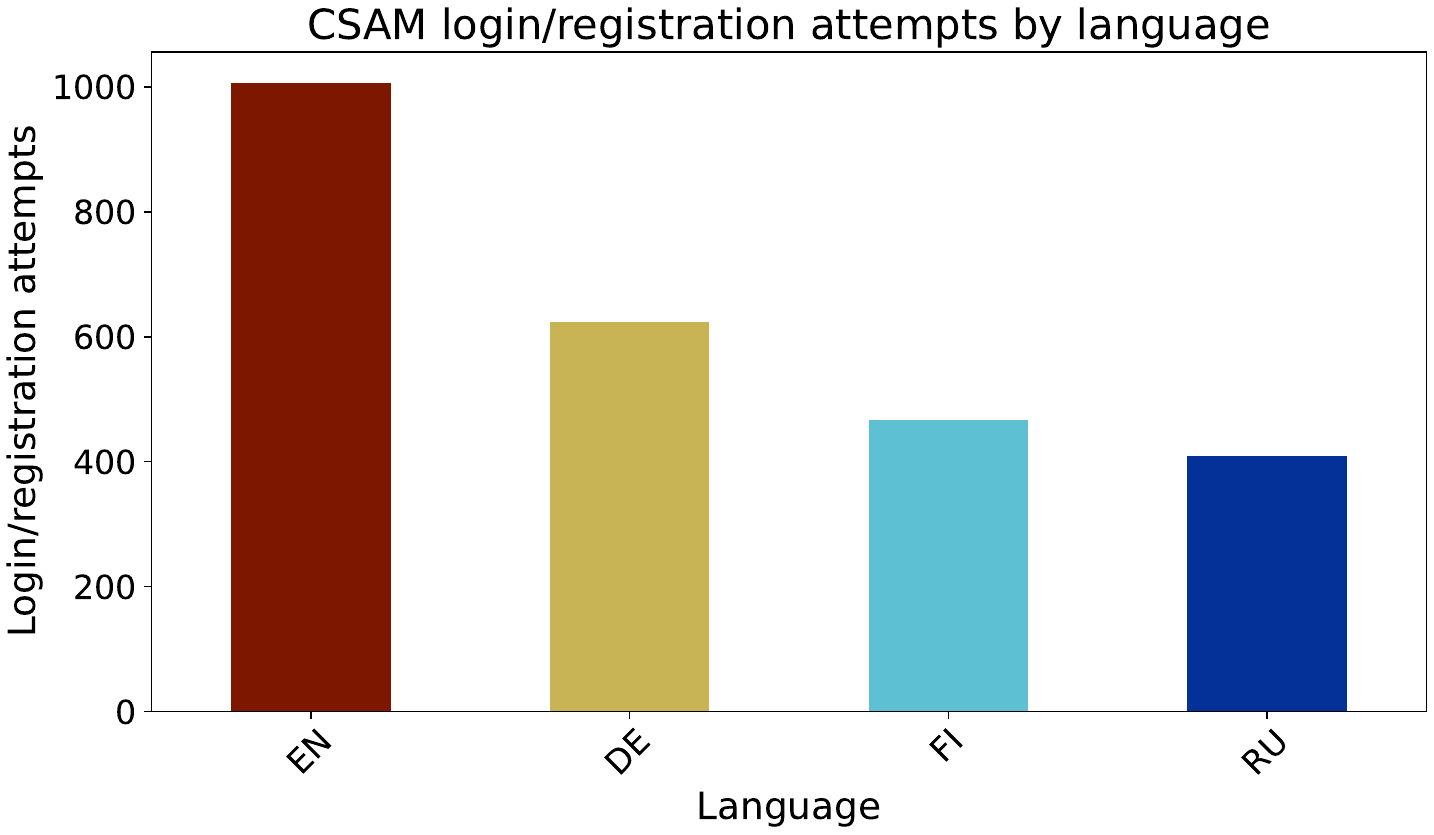}
    \caption{For the most popular category, CSAM, log in attempts per language.}\label{fig:csam_language}
\end{figure}

We also analyzed registration/log in attempts in the most popular category---\CSAM{}---for each language in~\autoref{fig:csam_language}.
The distribution was similar, with English being the most popular language with \rounding{1006/2499*100}\,\% (n = 1,006 of 2,499) of all log in attempts,
followed by German (\rounding{623/2499*100}\,\% or n = 623 of 2,499),
Finnish (\rounding{470/2499*100}\,\% or n = 470 of 2,499),
and Russian (\rounding{400/2499*100}\,\% or n = 400 of 2,499).
\textit{Our results clearly indicate that users prefer English over other languages.}

\section{Limitations}\label{sec:limitations}
\textbf{Our dataset captures only a subset of Tor activity:}
Not all Tor users access onion websites.
In 2015, the Tor Project estimated that onion service traffic accounts for about 3.4\% of total Tor traffic~\citep{dingledine2015onions}.
Moreover, Ahmia users represent only a subset of Tor search engine users, introducing sampling bias.

\textbf{\ahmia{} filters \CSAM{} from the search results and prevents CSAM-related queries:}
A major limitation is that human interactions in our dataset stem almost entirely from \ahmia{}, which blocks \CSAM{}-related queries.
Consequently, our RQ2 interpretations are conditional on a filtered, general-purpose search-engine discovery channel
and may not reflect behaviors of users who rely on unfiltered/private link-sharing or specialized forums.
Paradoxically, the high \CSAM{} engagement observed despite filtering is therefore surprising,
potentially reflecting a less-experienced audience arriving via search rather than seasoned offenders using alternative discovery paths.
Similarly, the low engagement with drug sites could be due to serious buyers already knowing the major marketplaces and having no reason to search for them on \ahmia{} or paste services.

\textbf{Honeypot realism:}
To allow for controlled comparisons, all of our honeypots had a minimalist and similar design.
This improves internal validity while decreasing realism in comparison to authentic illicit forums, potentially altering observed engagement tendencies.

\textbf{Language coverage:}
Our honeypots were available only in English, German, Finnish, and Russian.
Prior analyzes of Tor hidden-service content indicate that French and Italian also appear among the most common languages on onion websites.
Consequently, our dataset does not capture engagement from those linguistic communities.
Finnish was included primarily because we could generate accurate content easily, even though Finnish has a comparatively smaller presence on Tor than French and Italian.
Future work should expand honeypot deployments to include French and Italian to test whether the observed engagement patterns generalize across these additional languages.
\section{Conclusion}\label{sec:conclusion}

Methodologically, the honeypot design and the CAPTCHA test allowed us to distinguish between human and automated traffic.
We measured honeypot discovery and interaction across three channels:
the \ahmia{} Tor search engine, \pastebin{}, and \stronghold{}.
All three channels generated traffic, but human interactions originated almost entirely from \ahmia.

Analysis of registration/log in attempts---our proxy for genuine user interest---reveals that interaction events were disproportionately concentrated on abusive material.
The \CSAM{}-themed honeypot attracted most user engagement,
with more than twice the engagement of the next most popular category (violence).
Other categories with moderate interest included malware, stolen goods, and illegal firearms,
while illegal drugs and forgery received the least engagement.
Notably, the drug category's low engagement contrasts with its prominence in the larger dark web discourse.

In terms of language, English-language variants had the most interactions.
English was the language with both the highest overall and \CSAM{}-specific engagement,
followed by German and Finnish.
Despite its large global speaker base, the Russian language had the lowest user engagement.
Language preferences remained consistent across categories.
This overrepresentation of Finnish may be due to \ahmia{}'s origin and user base, rather than general dark web trends.
Overall, our findings reflect user engagement patterns that are specific to the discovery channel (public, general-purpose, filtered search engine) rather than the entire dark-web ecosystem.
 
\ifANON{}\else
\section*{Acknowledgment}
The authors declare no competing interests.
This work was supported by the European Commission under the Horizon Europe funding programme, as part of the project SafeHorizon (Grant Agreement 101168562). %
The content of this article does not reflect the official opinion of the European Union.
Responsibility for the information and views expressed therein lies entirely with the authors.
\fi

\ifIEEE{}
\bibliographystyle{IEEEtran}
\fi

\appendix\label{app:appendix}

\renewcommand{\sectionautorefname}{Appendix}
\renewcommand{\thesection}{\Alph{section}}
\ifANON\else
\section{Authorship contributions}\label{app:contributions}
Contributor Roles Taxonomy (\href{https://credit.niso.org/}{CRediT}) statement of authorship contribution.

\resizebox{0.8\linewidth}{!}{%
\scriptsize %
\begin{tabular}{
    p{0.35\linewidth}
    >{\centering\arraybackslash}p{0.08\linewidth}
    >{\centering\arraybackslash}p{0.08\linewidth}
    >{\centering\arraybackslash}p{0.08\linewidth}
    >{\centering\arraybackslash}p{0.08\linewidth}
}
\toprule
\textbf{Contributions} & \textbf{A.P.} & \textbf{W.A.} & \textbf{J.N.} \ \\
\midrule
Conceptualization & $\bullet$ & $\circ$ & $\bullet$ \\
Data curation & $\bullet$ & $\circ$ & $\bullet$ \\
Formal Analysis & $\bullet$ & $\bullet$ & $\bullet$ \\
Funding acquisition & $\circ$ & $\circ$ & $\bullet$ \\
Investigation & $\bullet$ & $\bullet$ & $\bullet$ \\
Methodology & $\bullet$ & $\circ$ & $\bullet$ \\
Project administration & $\bullet$ & $\circ$ & $\bullet$ \\
Resources & $\circ$ & $\circ$ & $\bullet$ \\
Software & $\circ$ & $\circ$ & $\bullet$ \\
Supervision & $\bullet$ & $\circ$ & $\bullet$ \\
Validation & $\bullet$ & $\bullet$ & $\bullet$ \\
Visualization & $\bullet$ & $\bullet$ & $\bullet$ \\
Writing---original draft & $\bullet$ & $\bullet$ & $\bullet$ \\
Writing---review \& editing & $\bullet$ & $\circ$ & $\bullet$ \\
\bottomrule
\end{tabular}%
}
 \fi

\section{Honeypot website screenshots}\label{app:screenshots}

\begin{figure}[!htb]
  \centering
    \includegraphics[width=\linewidth]{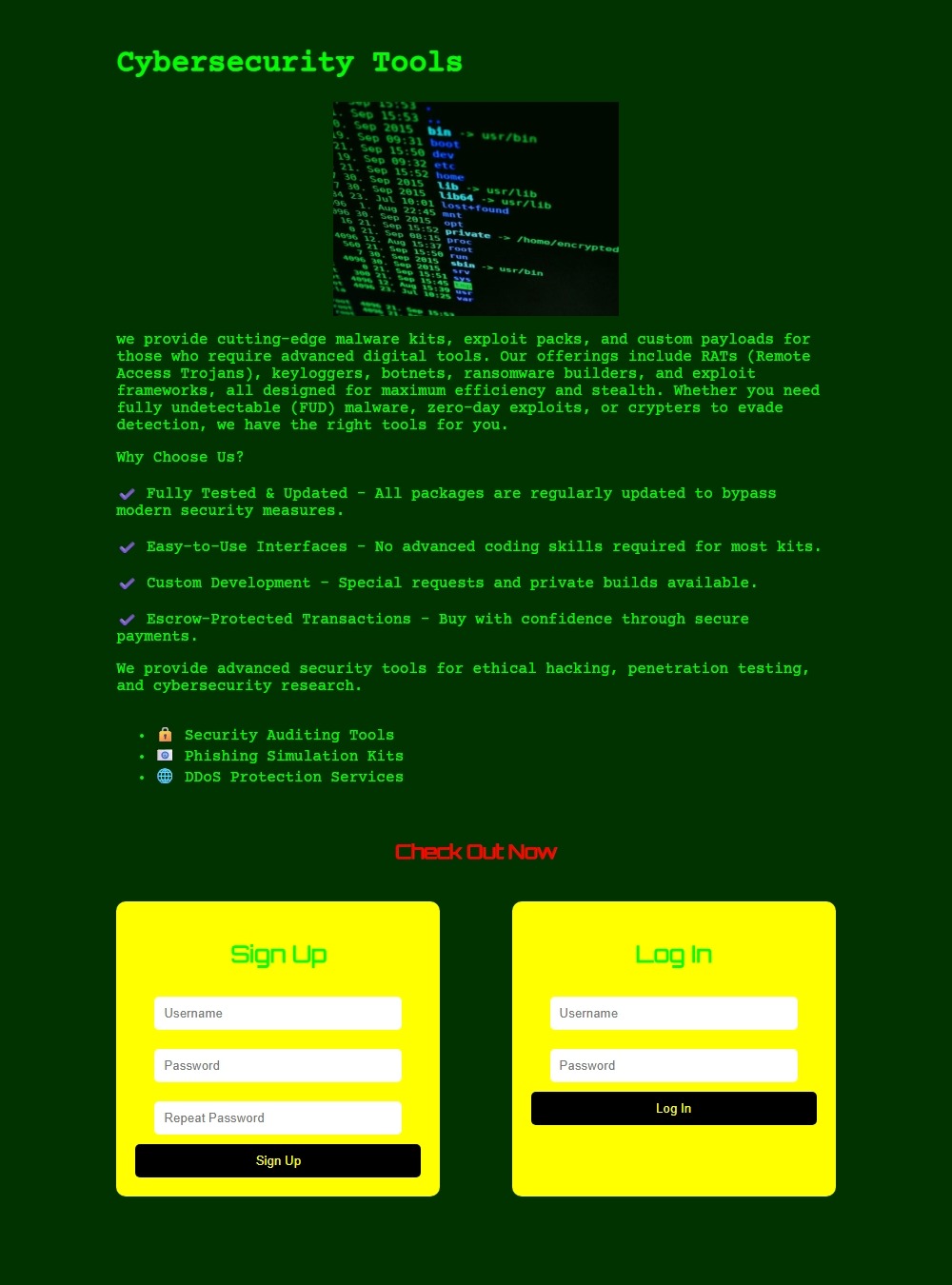}
    \caption{
    Screenshots of the honeypot websites (English versions): malware}\label{fig:malware}
\end{figure}

\begin{figure}[!htb]
  \centering
    \includegraphics[width=\linewidth]{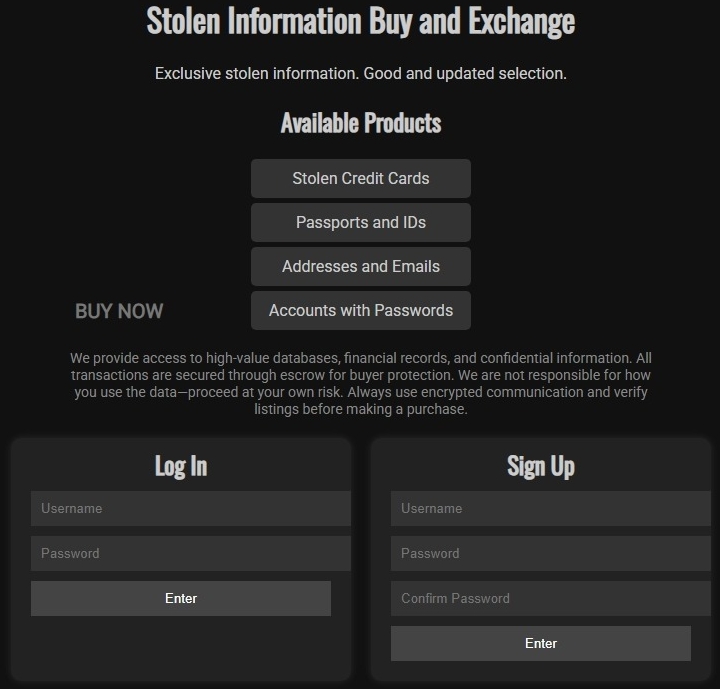}
    \caption{
    Stolen goods honeypot (English version)}\label{fig:stolen}
\end{figure}

\begin{figure}[!htb]
  \centering
    \includegraphics[width=\linewidth]{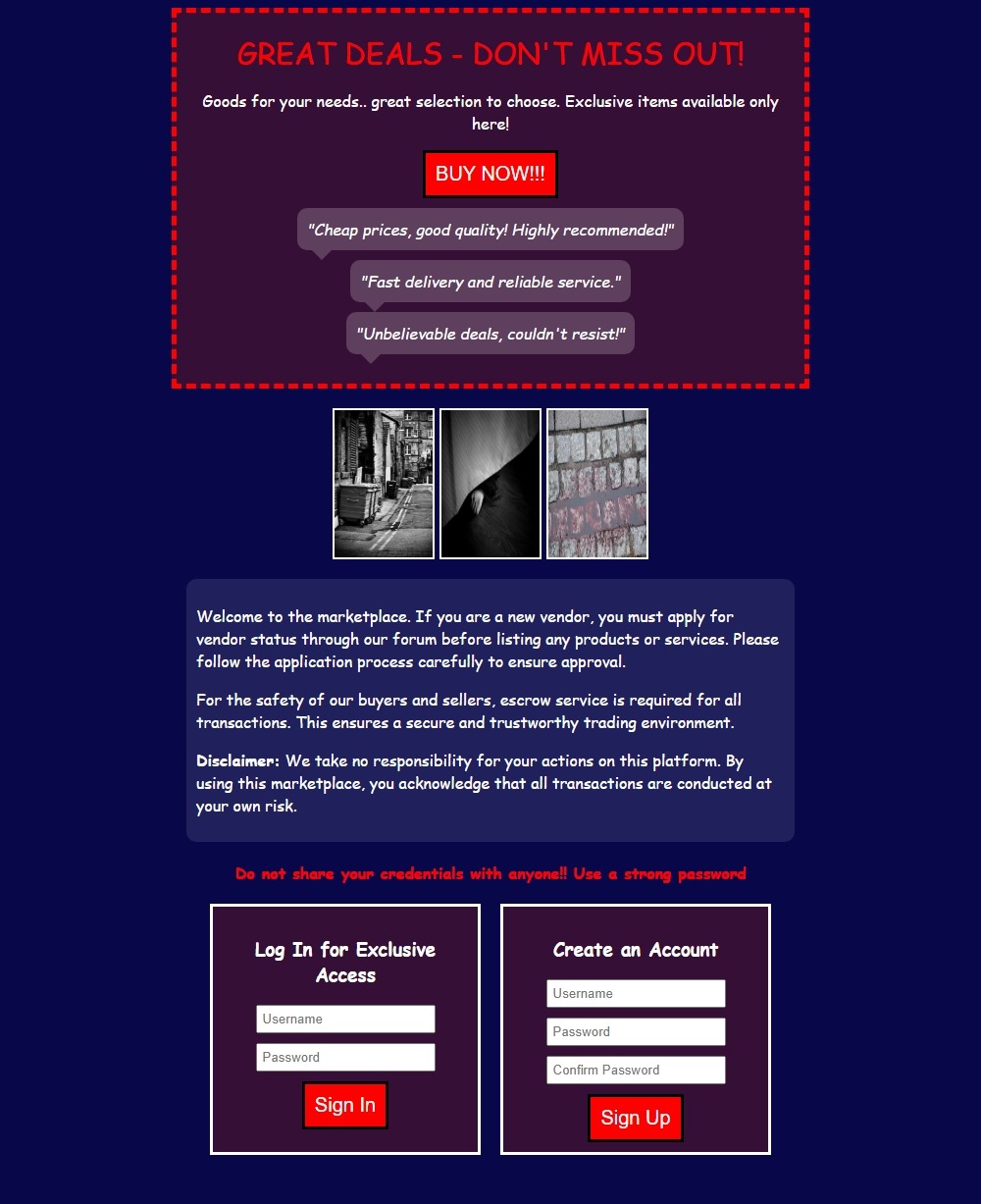}
    \caption{
    Unclear honeypot (English version)}\label{fig:unclear}
\end{figure}

\begin{figure}[!htb]
  \centering
    \includegraphics[width=\linewidth]{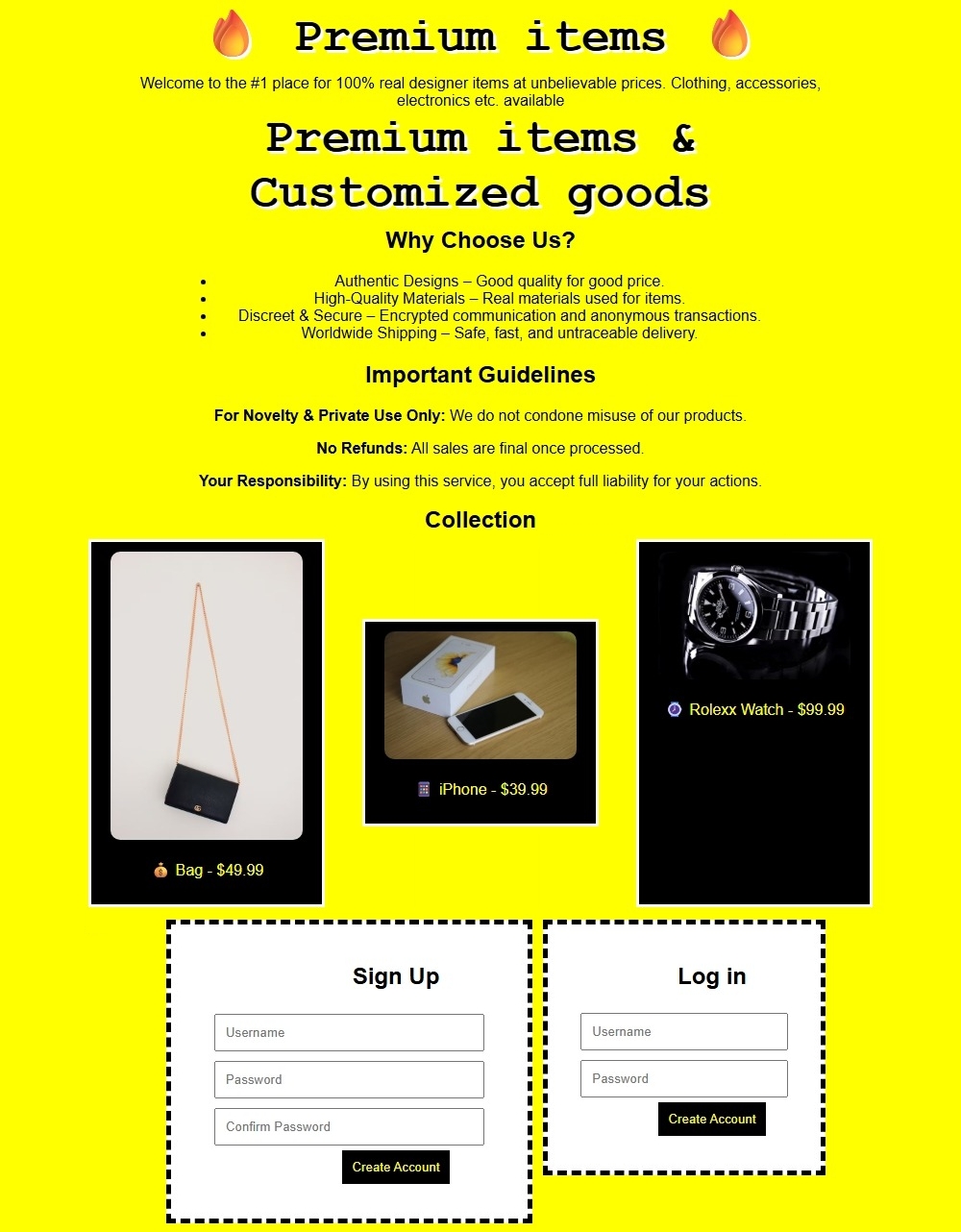}
    \caption{
    Forgery honeypot (English version)}\label{fig:forgery}
\end{figure}

\begin{figure}[!htb]
  \centering
    \includegraphics[width=\linewidth]{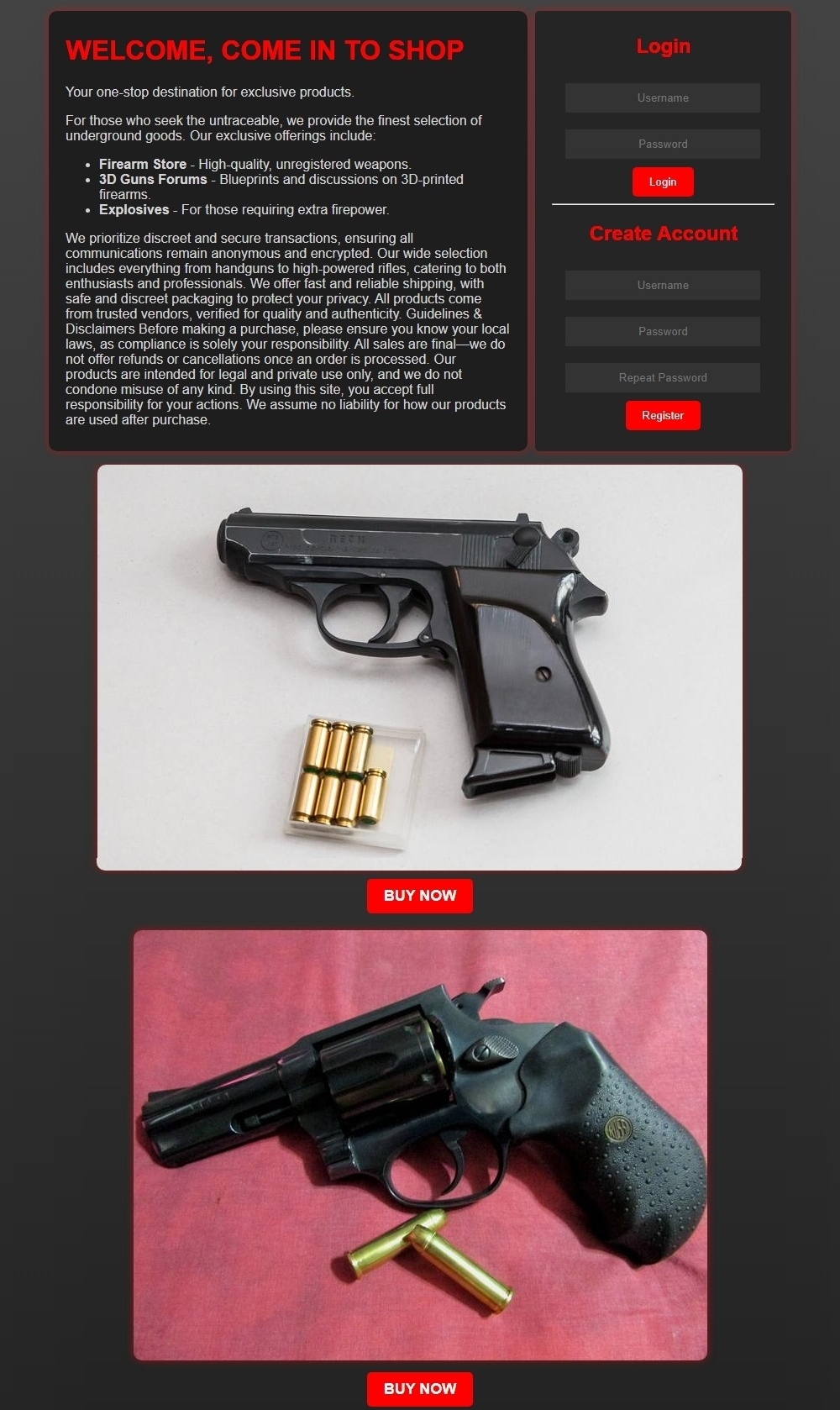}
    \caption{Firearms honeypot (English version)}\label{fig:firearms}
\end{figure}

\begin{figure}[!htb]
  \centering
    \includegraphics[width=\linewidth]{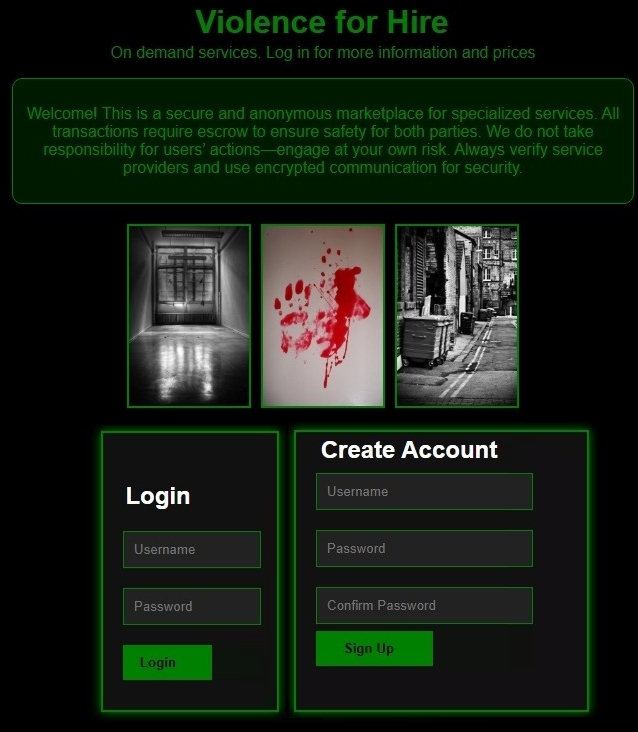}
    \caption{Violence honeypot (English version)}\label{fig:violence}
\end{figure}

\begin{figure}[!htb]
  \centering
    \includegraphics[width=\linewidth]{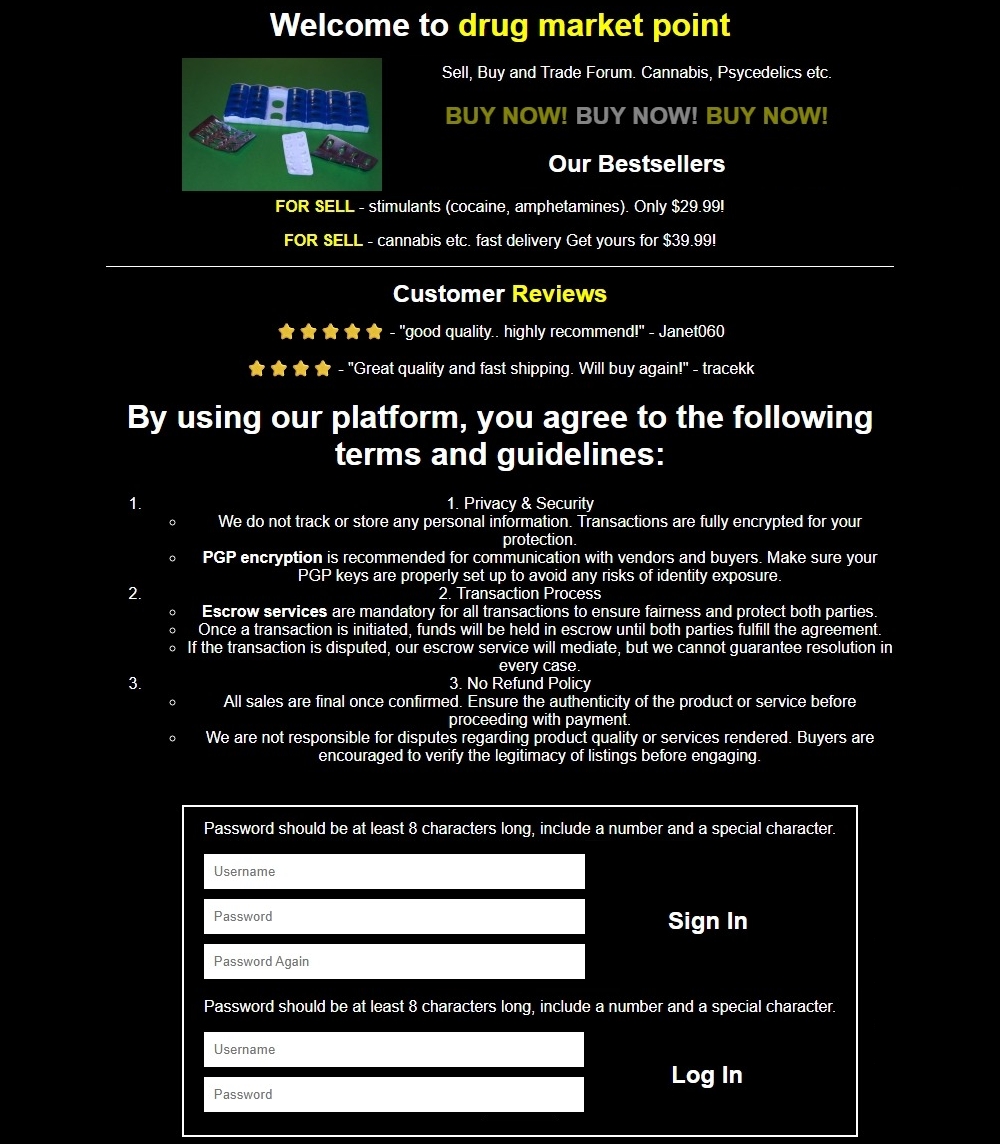}
    \caption{Drugs honeypot (English version)}\label{fig:drugs}
\end{figure}
 
\ifIACRTRANS{}
\printbibliography{}
\else

\fi

\end{document}